\input harvmac
\input amssym
\input epsf

\def\CN{{\cal N}}
\def\CE{{\cal E}}

\lref\AHJK{
  E.~Antonyan, J.~A.~Harvey, S.~Jensen and D.~Kutasov,
  ``NJL and QCD from string theory,''
[hep-th/0604017].
}
\lref\AK{
  O.~Aharony and D.~Kutasov,
  ``Holographic duals of long open strings,''
Phys.\ Rev.\ D {\bf 78} (2008) 026005.
[arXiv:0803.3547 [hep-th]].
}
\lref\ASY{
  O.~Aharony, J.~Sonnenschein and S.~Yankielowicz,
  ``A Holographic model of deconfinement and chiral symmetry restoration,''
Annals Phys.\  {\bf 322} (2007) 1420.
[hep-th/0604161].
}
\lref\BLL{
  O.~Bergman, G.~Lifschytz and M.~Lippert,
  ``Holographic nuclear physics,''
JHEP {\bf 0711} (2007) 056.
[arXiv:0708.0326 [hep-th]].
}
\lref\BR{
  G.~E.~Brown and M.~Rho,
  ``Scaling effective Lagrangians in a dense medium,''
Phys.\ Rev.\ Lett.\  {\bf 66} (1991) 2720.
}
\lref\BSS{
  O.~Bergman, S.~Seki and J.~Sonnenschein,
  ``Quark mass and condensate in HQCD,''
JHEP {\bf 0712} (2007) 037.
[arXiv:0708.2839 [hep-th]].
}
\lref\CKP{
  R.~Casero, E.~Kiritsis and A.~Paredes,
  ``Chiral symmetry breaking as open string tachyon condensation,''
Nucl.\ Phys.\ B {\bf 787} (2007) 98.
[hep-th/0702155 [HEP-TH]].
}
\lref\CPS{
  R.~Casero, A.~Paredes and J.~Sonnenschein,
  ``Fundamental matter, meson spectroscopy and non-critical string/gauge duality,''
JHEP {\bf 0601} (2006) 127.
[hep-th/0510110].
}
\lref\DN{
  A.~Dhar and P.~Nag,
  ``Tachyon condensation and quark mass in modified Sakai-Sugimoto model,''
Phys.\ Rev.\ D {\bf 78} (2008) 066021.
[arXiv:0804.4807 [hep-th]].
}
\lref\EV{
  M.~Edalati and J.~F.~Vazquez-Poritz,
  ``Chiral condensates in finite density holographic NJL model from string worldsheets,''
[arXiv:0906.5336 [hep-th]].
}
\lref\HHLY{
  K.~Hashimoto, T.~Hirayama, F.~-L.~Lin and H.~-U.~Yee,
  ``Quark mass deformation of holographic massless QCD,''
JHEP {\bf 0807} (2008) 089.
[arXiv:0803.4192 [hep-th]].
}
\lref\HHM{
  K.~Hashimoto, T.~Hirayama and A.~Miwa,
  ``Holographic QCD and pion mass,''
JHEP {\bf 0706} (2007) 020.
[hep-th/0703024 [HEP-TH]].
}
\lref\HSSY{
  H.~Hata, T.~Sakai, S.~Sugimoto and S.~Yamato,
  ``Baryons from instantons in holographic QCD,''
Prog.\ Theor.\ Phys.\  {\bf 117} (2007) 1157.
[hep-th/0701280 [HEP-TH]].
}
\lref\KSZ{
  K.~-Y.~Kim, S.~-J.~Sin and I.~Zahed,
  ``Dense hadronic matter in holographic QCD,''
[hep-th/0608046].
}
\lref\KSZi{
  K.~-Y.~Kim, S.~-J.~Sin and I.~Zahed,
  ``The chiral model of Sakai-Sugimoto at finite baryon density,''
JHEP {\bf 0801} (2008) 002.
[arXiv:0708.1469 [hep-th]].
}
\lref\KSZii{
  K.~-Y.~Kim, S.~-J.~Sin and I.~Zahed,
  ``Dense holographic QCD in the Wigner-Seitz approximation,''
JHEP {\bf 0809} (2008) 001.
[arXiv:0712.1582 [hep-th]].
}
\lref\Ma{
  J.~M.~Maldacena,
  ``The Large N limit of superconformal field theories and supergravity,''
Adv.\ Theor.\ Math.\ Phys.\  {\bf 2} (1998) 231, [Int.\ J.\ Theor.\ Phys.\  {\bf 38} (1999) 1113].
[hep-th/9711200].
}
\lref\Mai{
  J.~M.~Maldacena,
  ``Wilson loops in large N field theories,''
Phys.\ Rev.\ Lett.\  {\bf 80} (1998) 4859.
[hep-th/9803002].
}
\lref\NSSY{
  S.~Nakamura, Y.~Seo, S.~-J.~Sin and K.~P.~Yogendran,
  ``A new phase at finite quark density from AdS/CFT,''
J.\ Korean Phys.\ Soc.\  {\bf 52} (2008) 1734.
[hep-th/0611021].
}
\lref\PSM{
  K.~Peeters, J.~Sonnenschein and M.~Zamaklar,
  ``Holographic melting and related properties of mesons in a quark gluon plasma,''
Phys.\ Rev.\ D {\bf 74} (2006) 106008.
[hep-th/0606195].
}
\lref\RSZ{
  M.~Rho, S.~-J.~Sin and I.~Zahed,
  ``Dense QCD: A holographic dyonic salt,''
Phys.\ Lett.\ B {\bf 689} (2010) 23.
[arXiv:0910.3774 [hep-th]].
}
\lref\SaSu{
  T.~Sakai and S.~Sugimoto,
  ``Low energy hadron physics in holographic QCD,''
Prog.\ Theor.\ Phys.\  {\bf 113} (2005) 843.
[hep-th/0412141].
}
\lref\SaSui{
  T.~Sakai and S.~Sugimoto,
  ``More on a holographic dual of QCD,''
Prog.\ Theor.\ Phys.\  {\bf 114} (2005) 1083.
[hep-th/0507073].
}
\lref\Se{
  S.~Seki,
  ``Intersecting D4-branes model of holographic QCD and tachyon condensation,''
JHEP {\bf 1007} (2010) 091.
[arXiv:1003.2971 [hep-th]].
}
\lref\SeSi{
  Y.~Seo and S.~-J.~Sin,
  ``Baryon mass in medium with holographic QCD,''
JHEP {\bf 0804} (2008) 010.
[arXiv:0802.0568 [hep-th]].
}
\lref\SeSo{
  S.~Seki and J.~Sonnenschein,
  ``Comments on baryons in holographic QCD,''
JHEP {\bf 0901} (2009) 053.
[arXiv:0810.1633 [hep-th]].
}
\lref\Wit{
  E.~Witten,
  ``Baryons and branes in anti-de Sitter space,''
JHEP {\bf 9807} (1998) 006.
[hep-th/9805112].
}

\Title{}
{\vbox{\centerline{Chiral Condensate in Holographic QCD}
	\vskip12pt\centerline{with Baryon Density}}}
 
\centerline{Shigenori Seki\footnote{$^\dagger$}
{\tt sigenori@apctp.org}}
\medskip\centerline{\it Asia Pacific Center for Theoretical Physics}
\centerline{\it San 31, Hyoja-dong, Nam-gu, Pohang, Gyeongbuk 790-784, Republic of Korea}
\bigskip
\centerline{Sang-Jin Sin\footnote{$^\ddagger$}
{\tt sjsin@hanyang.ac.kr}}
\medskip\centerline{\it Department of Physics, Hanyang University, Seoul 133-791, Republic of Korea}
 
\vskip .3in 
 
\centerline{\bf Abstract} 

We consider the chiral condensate 
in the baryonic dense medium using the generalized Sakai-Sugimoto model.
It is defined as the vacuum expectation value of open Wilson line 
that is proposed to be calculated by use of the area of world-sheet instanton. 
We evaluate it in confined as well as deconfined phase. 
In both phases, the chiral condensate has a minimum 
as a function of  baryon density. 
In the deconfined phase, taking into account the chiral symmetry restoration, 
we classify the behavior of chiral condensate into three types. 
One can set the parameter of the theory such that the results, 
in low but sufficiently higher density, 
is in agreement with the expectation from QCD. 

\Date{26 June 2012}

\newsec{Introduction}

One of the most important problem in strong interaction physics is 
the understanding of the chiral symmetry and 
calculation of its order parameter in dense medium, 
since it is the key quantity determining the hadron property 
in the  nuclei as well as in neutron stars. 
For long time, there has been many speculation on the behaviour 
of chiral condensation in dense medium: 
one of the intuitively compelling one is that as an order parameter, 
it is non-zero at vacuum and vanishes at the large enough density 
where the chiral symmetry is restored. 
Therefore the most naive but natural behaviour is to decrease 
as a function of density until it vanishes at the transition point \BR.  
However, to our knowledge, there is no work which prove  
this scenario  from the first principle. 
This is because the strongly interacting nature and 
the presence of the chemical potential have been blocking any reliable calculation 
even in the numerical approach: 
the Dyson-Schwinger equation for the resummation  is not justified 
in strong coupling since it involve the truncation of significant part of the diagrams. Furthermore 
the solution of notorious sign problem in lattice is not available yet. 
 
Since this situation, there have been many activities 
to utilize the gauge/gravity correspondence \Ma\ to improve 
the understanding of the nuclear physics. 
Most notable one is the model of Sakai-Sugimoto \refs{\SaSu,\SaSui}, 
since the chiral symmetry and its breaking is intuitively 
realized in geometric way.  
However, it has its own drawback, 
since one can not include the quark mass in natural way 
and calculating chiral condensation is blocked for similar reason. 
So far, in order to improve this problem, many prescriptions 
({\it e.g.}~the uses of tachyonic DBI action \refs{\CKP\BSS\DN{--}\Se}, 
the introduction of additional D4-branes or D6-branes \refs{\HHM,\HHLY}, 
{\it etc.}) have been suggested.
Especially in this paper we are interested in the idea 
suggested by Aharony and  Kutasov \AK\ to utilize 
the world-sheet instanton for quark mass and chiral condensation. 
The idea is to separate the ${\bar \psi}$ and $\psi$ 
such that one is in D8-branes and the other is in anti-D8-branes 
in the UV region 
and covariantize the composite operator 
by inserting the open Wilson line connecting them. 
Then the gravity dual of this operator 
is the world-sheet instanton, a minimal area whose boundary is given by the open Wilson line connecting 
$\bar \psi$ and $\psi$. 
This is analogous to the Wilson loop calculation in the closed Wilson line case \Mai. 

The baryons in the Sakai-Sugimoto model are realised as
the instantons, in other words, Skyrme-like solitons from 
the viewpoint of the flavour D8-branes \HSSY, 
while the baryons are also interpreted to the baryon vertex 
given by the D4-branes wrapping $S^4$, which we call baryonic D4-branes \Wit. 
Therefore one can describe the baryonic medium 
by the use of the baryonic D4-branes which are 
uniformly distributed in $\Bbb{R}^3$ \BLL.
The Sakai-Sugimoto model in dense medium 
has been attracting many interests \refs{\BLL\KSZ\KSZi{--}\KSZii} 
as well as the other holographic models have \refs{\NSSY,\SeSi}. 
In a recent paper, the authors of Ref.~\EV\ worked out the problem of 
chiral symmetry order parameter in baryonic matter 
in the context of holographic NJL model \AHJK, 
which is uncompatified version of Sakai-Sugimoto model, 
where there is no confinement.  
In this paper we will work out the calculation in compactified version 
with confinement. 

We will find that we can tune the parameters of the Sakai-Sugimoto model 
such that chiral condensation decreases as function of density 
until density reach the chiral transition point, 
where it suddenly drops to zero. 
That is, it does not decreases to zero continuously 
but will go to the first order transition. 

The remaining of the paper goes as follows: 
In Section 2, we review the definition of 
the generalised Sakai-Sugimoto model \ASY\
and the open Wilson line operator as chiral condensate \AK. 
In Section 3, we analyse the confined phase. 
After showing the force balance condition by following Ref.~\BLL, 
we numerically calculate the chiral condensate, 
and compare it with the normal nuclear density 
by the use of the experimental values.
In Section 4, in the way similar to the confined phase, 
we investigate the chiral condensate and the chiral symmetry restoration 
in the deconfined phase. 
Section 5 is devoted for the conclusion and discussion.

\newsec{Chiral condensate in generalised Sakai-Sugimoto model}

\subsec{Generalised Sakai-Sugimoto model}

The Sakai-Sugimoto model \refs{\SaSu, \SaSui} consists of $N_c$ colour D4-branes 
and $N_f$ pairs of flavour D8-branes and anti-D8-branes. 
The open strings connecting the D4-branes and the D8-branes (anti-D8-branes) 
describe quarks (anti-quarks). 
In large $N_c$ where the AdS/CFT correspondence is valid, 
we can interpret the colour D4-branes as a background which has two phases. 
\bigskip\vbox{
\centerline{\epsfbox{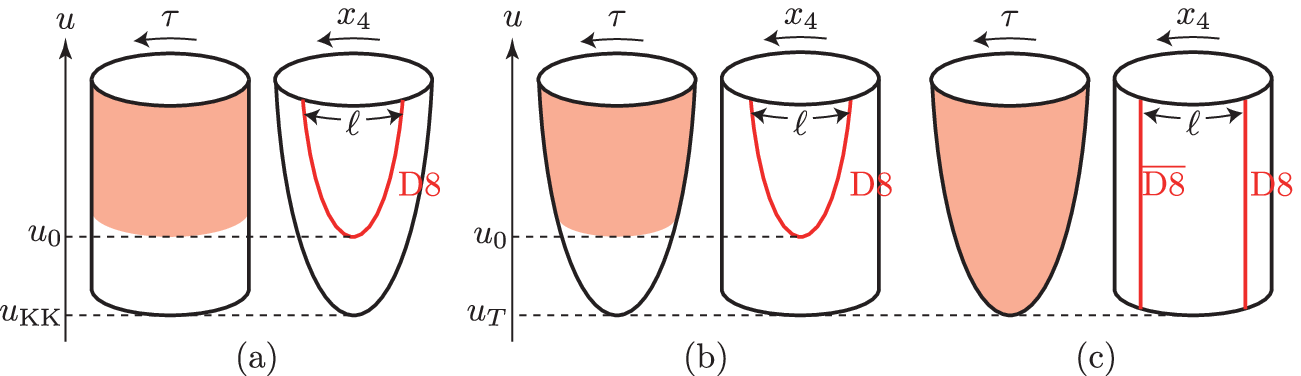}} 
\centerline{\vbox{\offinterlineskip 
\halign{ \strut# & #\hfil \cr
\fig\figphases{} & (a) The confined phase where the chiral symmetry is broken. \cr
 &  (b) The deconfined phase where the chiral symmetry is broken. \cr
 &	(c) The deconfined phase where the chiral symmetry is restored. \cr
}}}
}\bigskip\nobreak\noindent
One is the confined phase (\figphases a), 
the other is the deconfined phase (\figphases b,c). 
The transition between these two phases is realised as Hawking-Page transition. 
We can deal with the flavour D8-branes and anti-D8-branes as a probe 
in this background. 

The background (Euclidean) metric in the confined phase is described as 
\eqna\Cmetric
$$\eqalignno{
&R^{-2}ds^2 = u^{3 \over 2} \biggl[d\tau^2 +\sum_{i=1}^3 (dx_i)^2 +f(u) (dx_4)^2\biggr] 
	+u^{-{3 \over 2}}\biggl[{du^2 \over f(u)} +u^2 d\Omega_4^2 \biggr] \,, &\Cmetric{a} \cr
&e^\phi = g_s u^{3 \over 4} \,, \quad 
F_4 = {(2\pi)^3 (\alpha')^{3 \over 2} N_c \over \Omega_4} \epsilon_4 \,, \quad 
R^3 =\pi g_s N_c (\alpha')^{3/2} \,, &\Cmetric{b} \cr
&f(u) = 1 - \biggl({u_{\rm KK} \over u}\biggr)^3 \,. &\Cmetric{c}
}$$
Note that the coordinates are dimensionless due to rescaling 
by the radius $R$ which has a dimension of length. 
$\Omega_4$ is the volume of a unit $S^4$, {\it i.e.}, $\Omega_4 = 8\pi^2/3$. 
The D4-branes are originally spanned 
in the $(t, x_1, x_2, x_3, x_4)$ directions 
where $t$ is Wick-rotated as $t \to i \tau$. 
The $x_4$ direction is compact with a period $\beta_4$, 
namely, $x_4 \sim x_4 + \beta_4$. 
Although generally this yields a conical singularity at $u = u_{\rm KK}$, 
one can remove it by choosing 
\eqn\Cfourper{
\beta_4 = {4 \pi \over 3} {1 \over \sqrt{u_{\rm KK}}} \,,
}
so that the $u$-$x_4$ space has cigar geometry (see \figphases a). 
The period $\beta_4$ leads to the Kaluza-Klein Mass and the Yang-Mills coupling, 
$$M_{\rm KK} = {2 \pi \over R \beta_4} \,, \quad 
g_{\rm YM}^2 = {(2\pi)^2 g_s l_s \over R\beta_4} \,.
$$
Therefore one can obtain the relations between 
the parameters in the gravity side, $R, u_{\rm KK}$ and $g_s$, 
and those in the Yang-Mills theory side, 
$M_{\rm KK}$ and $\lambda \,(:=g_{\rm YM}^2 N_c)$, as follows:
\eqn\relGrYM{
R^3 = {\lambda l_s^2 \over 2 M_{\rm KK}} \,, \quad 
u_{\rm KK} = {2^{4 \over 3} \over 9}
	(\lambda M_{\rm KK}^2 l_s^2)^{2 \over 3} \,, \quad 
g_s = {\lambda \over 2\pi N_c M_{\rm KK} l_s} \,.
}
In order to introduce a temperature in this phase, 
we compactify the $\tau$ direction with a period $\beta_\tau$, 
so that the temperature is given by $2\pi/(R\beta_\tau)$. 

In the deconfined phase (\figphases b,c), the background metric is 
\eqna\Dmetric
$$\eqalignno{
&R^{-2}ds^2 = u^{3 \over 2} \biggl[ f_T(u) d\tau^2 +\sum_{i=1}^3 (dx_i)^2 +(dx_4)^2 \biggr] 
	+u^{-{3 \over 2}}\biggl[ {du^2 \over f_T(u)} +u^2 d\Omega_4^2 \biggr] \,, &\Dmetric{a} \cr
&e^\phi = g_s u^{3 \over 4} \,, \quad F_4 = {(2\pi)^3 (\alpha')^{3 \over 2} N_c \over \Omega_4} \epsilon_4 \,, \quad 
R^3 =\pi g_s N_c (\alpha')^{3 \over 2} \,, &\Dmetric{b} \cr
&f_T(u) = 1 - \biggl({u_T \over u}\biggr)^3 \,. &\Dmetric{c}
}$$
This Euclidean metric allows us to consider thermodynamics in the deconfined phase. 
The $\tau$ direction is compactified with a period $\beta_\tau$, 
{\it i.e.}, $\tau \sim \tau + \beta_\tau$, 
so that the period should be related to $u_T$ as 
\eqn\Dtper{
\beta_\tau = {4\pi \over 3} {1 \over \sqrt{u_T}} \,,
}
by the same reason as \Cfourper. 
Then the Hawking temperature $T$ 
is denoted by $T = 2\pi/(R \beta_\tau)$. 
By this compactification the $\tau$-$u$ space 
of the background becomes a cigar (\figphases b,c).  
We also compactify the $x_4$ direction with the period $\beta_4$. 

In the backgrounds, \Cmetric{} and \Dmetric{}, 
we consider the flavour D8-branes and anti-D8-branes as a probe. 
In the confined phase, the D8-branes and anti-D8-branes are connected 
and become U-shape D8-branes. 
Since the pair of D8-branes and anti-D8-branes describe 
$U(N_f)_L \times U(N_f)_R$ chiral symmetry, the U-shape D8-branes 
imply the chiral symmetry breaking, $U(N_f)_L \times U(N_f)_R \to U(N_f)$.
On the other hand, there are two possibilities in the deconfined phase. 
One is the chiral symmetry broken phase which is represented 
by the U-shape D8-branes (\figphases b). 
The other is the chiral symmetry restored phase given by 
the configuration of parallel D8-branes and anti-D8-branes (\figphases c).

In the original Sakai-Sugimoto model, the D8-branes and anti-D8-branes 
are placed at the antipodal position in the circle of $x_4$ direction, 
in other words, the separation $\ell$ between the D8-branes 
at the UV limit, $u=\infty$, 
is a half of the period of $x_4$ direction, $\ell = \beta_4/2$. 
However one can generalise the separation to be any $\ell$ less than $\beta_4/2$. 
In this paper we consider this generalised Sakai-Sugimoto model, 
to which Bergman {\it et al.}~introduced a baryon density \BLL. 

\bigskip\vbox{
\centerline{\epsfbox{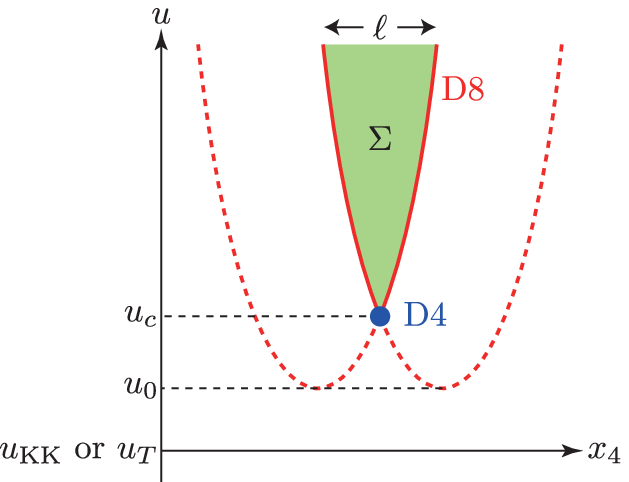}} 
\centerline{\vbox{\offinterlineskip 
\halign{ \strut# & #\hfil \cr
\fig\figVshape{} & The V-shape D8-branes with baryonic D4-branes. The green shaded region \cr 
	& is the world-sheet of string corresponding to the open Wilson line. \cr}}}
}\bigskip\nobreak\noindent

The baryons are realised as the D4-branes wrapping $S^4$ which are located 
at the tip of the flavour D8-branes. 
Therefore the system of generalised Sakai-Sugimoto model with baryons 
should obey the balance condition for 
the forces between the D4-branes and the D8-branes. 
This condition gives rise to the modification of the D8-branes 
from U-shape to V-shape (\figVshape) \BLL.

\subsec{Open Wilson line as chiral condensate}

So far many trials for incorporating a chiral condensate 
into the Sakai-Sugimoto model have been done. 
In this paper we adopt the method introduced by Aharony and Kutasov \AK. 
Let us consider the open Wilson line operator, 
$$
\CO^j_i(x^\mu) = \psi_L^{\dagger j}\biggl(x^\mu,x^4=-{\ell \over 2}\biggr) 
{\cal P}\exp\Biggl[\int_{-\ell/2}^{\ell/2}(iA_4 +\Phi)dx^4\Biggr] \psi_{Ri}\biggl(x^\mu,x^4={\ell \over 2}\biggr) \,,
$$
where $\Phi$ is one of the scalar fields in the super-Yang-Mills theory 
and $i,j$ are indices of the fundamental representation of $U(N_f)$. 
This corresponds to the well-known order parameter of 
chiral symmetry breaking, $\psi_L^{\dagger j}\psi_{Ri}$, in QCD. 
In the D-brane configuration that we are considering, the operator $\psi$ is 
at the UV boundary on the V-shape (or U-shape) D8-branes. Therefore, 
following the AdS/CFT correspondence, 
one can calculate the vacuum expectation value of $\CO^j_i$, 
$$
\langle \CO^j_i \rangle \simeq \delta_{ij} \langle \CO \rangle \,, \quad 
\langle \CO \rangle = e^{-S_\CO} \,, 
$$
where $S_\CO$ is the on-shell Euclidean action of the open string 
whose world-sheet is bounded by the flavour D8-branes (see \figVshape). 
To leading order in $\alpha'$, $S_\CO$ is calculated 
by the minimal area of the string world-sheet, 
\eqn\strWS{
S_{\CO} = {1 \over 2\pi\alpha'}\int_\Sigma d^2\sigma\, \sqrt{\det g} \,, 
}
where $\Sigma$ is the region surrounded by the flavour D8-branes (see \figVshape). 
$\langle \CO \rangle$ is proportional to $m_q \langle {\bar \psi}\psi \rangle$, 
and indeed Ref.~\AK\ has shown that $\langle \CO \rangle$ naively satisfies 
the Gell-Mann-Oakes-Renner relation, $m_\pi^2 f_\pi^2 \propto \langle \CO \rangle$.

\newsec{Confined phase}

\subsec{D-brane configuration and force balance condition}

In this subsection, following Ref.~\BLL, we shall show 
the V-shape solution of flavour D8-branes, the baryonic D4-branes wrapping $S^4$ 
and their force balance condition. 

We start with the action of the $N_f$ flavour D8-branes 
embedded in the background \Cmetric{} 
which have the world-volume coordinates $(\tau,x_1,x_2,x_3,u,\Omega_4)$ 
and the collective coordinate $x_4(u)$. 
The world-volume gauge fields on the D8-branes are decomposed into 
$$
\CA = A_{SU(N_f)} + {1 \over \sqrt{2N_f}}A_{U(1)} \,. 
$$
Since the $U(1)$ part is related with the baryon density 
that we are here interested in, 
we turn on 
\eqn\uoneat{
a_\tau(u) := {2\pi\alpha' \over R^2\sqrt{2N_f}}A_{U(1),\tau}(u) \,,
}
where we assumed that the gauge field depends only on $u$. 
Note that the rescaled gauge field $a_\tau(u)$ is dimensionless. 
Then the DBI action of D8-branes is written down as 
\eqna\CDeightDBI
$$\eqalignno{
&S_8^{\rm (DBI)} = -\CN V_3 \int d\tau du\, \CL[x_4',a_\tau'] \,, &\CDeightDBI{a} \cr
&\CL[x_4',a_\tau'] = 
	u^4 \sqrt{ f(u) \bigl(x_4'(u)\bigr)^2 
	-{1 \over u^3}\bigl(a_\tau'(u)\bigr)^2 +{1 \over u^3 f(u)} } \,, &\CDeightDBI{b}
}$$
where $\CN = N_f \mu_8 \Omega_4 R^9 / g_s$ 
and $V_3 (= \int d^3x)$ is the volume of $\Bbb{R}^3$. 
$\mu_8$ is the D8-brane's tension, 
{\it i.e.}, $\mu_8 = (2\pi)^{-8}(\alpha')^{-9/2}$. 
Note that $V_3$ is a  dimensionless value because the coordinates 
of $\Bbb{R}^3$, $x_i$ ($i=1,2,3$), are dimensionless. 
We introduce an electric displacement field $d(u)$ by 
\eqn\Celedis{
d(u) := -{\delta \CL \over \delta a_\tau'(u)} 
= u a_\tau'(u) \biggl[ f(u) \bigl(x_4'(u)\bigr)^2 -{1 \over u^3} \bigl(a_\tau'(u)\bigr)^2 +{1 \over u^3 f(u)} \biggr]^{-{1 \over 2}} \,.
}
This will later be associated with the baryon density. 
In terms of $d(u)$, the equations of motion for $x_4(u)$ and $a_\tau(u)$ 
from the action \CDeightDBI{} are expressed as 
\eqna\Ceom
$$\eqalignno{
0&= {d \over du}\Biggl[u^4 f(u) x_4'(u) 
	\biggl(1 + {\bigl(d(u)\bigr)^2 \over u^5}\biggr)^{1 \over 2} 
	\biggl( f(u) \bigl(x_4'(u)\bigr)^2 +{1 \over u^3 f(u)} \biggr)^{-{1 \over 2}} 
\Biggr] \,, &\Ceom{a} \cr
0&=d'(u) \,. &\Ceom{b} \cr
}$$
Integrating these equations once with respect to $u$, 
we obtain 
\eqna\CDeightsol
$$\eqalignno{
\bigl(x_4'(u)\bigr)^2 &= {1 \over u^3 \bigl(f(u)\bigr)^2}\biggl[ 
	{f(u)(u^8 +u^3 d^2) \over f(u_0)(u_0^8 +u_0^3d^2)} -1 \biggr]^{-1} \,, &\CDeightsol{a} \cr
d(u) &= d \,, &\CDeightsol{b}
}$$
where $d$ and $u_0$ are integration constants. 
$u_0$ is the point at which $x_4'$ diverges (see \figVshape).

The Chern-Simons action of the D8-branes, 
$$
S_8^{\rm (CS)} = {\mu_8 \over 3!} \int C_3 \wedge \Tr(2\pi\alpha' \CF')^3 
= {N_c \over 24\pi^2} \int \omega_5(\CA) \,,
$$
where $\omega_5(\CA) = \Tr(\CA \CF^2 -2^{-1}\CA^3 \CF +10^{-1}\CA^5)$, 
introduces a source of $a_\tau(u)$ at $u=u_c\, (\geq u_0)$:
$$
N_c V_3 \int d\tau du\, {1 \over \sqrt{2N_f}}A_{U(1),\,\tau} {1 \over 8\pi^2}\tr F_{SU(N_f)}^2 \,.
$$
We assume a uniform distribution of baryons 
in the $\Bbb{R}^3$ of $x^i$ $(i=1,2,3)$, namely, 
\eqn\fourchg{
{1 \over 8\pi^2}\tr F_{SU(N_f)}^2 = n_4 \delta(u-u_c) \,.
}
$n_4$ implies a baryon density and indeed 
one can measure the realistic baryon number density $n_B$ 
with $({\rm length})^{-3}$ dimension by 
\eqn\BNdensity{
n_B := {n_4 \over R^3} \,.
}
$n_4$ also corresponds to a dimensionless density parameter 
of baryonic D4-branes wrapping $S^4$. 
By \uoneat, the source term can be written as 
$$
S_8^{\rm (source)} 
= {N_c R^2 V_3 \over 2\pi\alpha'}\int d\tau  du\, a_\tau(u) n_4 \delta(u-u_c) \,.
$$
Therefore the equation of motion for $a_\tau(u)$ 
coming from the total action, $S_8^{\rm DBI} + S_8^{\rm (source)}$, is 
$$
\CN d'(u) = {N_c R^2 \over 2\pi\alpha'} n_4 \delta(u-u_c) \,. 
$$
By integrating this with respect to $u$, 
the densities, $d$ and $n_4$, are related by 
\eqn\dnfour{
d = {N_c R^2 \over 2\pi\alpha' \CN}n_4 \,.
}

The source of baryon charge can be interpreted as $N_4 (= n_4 V_3)$ 
baryonic D4-branes wrapping $S^4$ at $u=u_c$. 
Since the DBI action of the D4-branes is 
$$
S_4 = -N_4 \mu_4 \int d\tau d\Omega_4\, e^{-\phi}\sqrt{\det g} \,, 
$$
where $\mu_4 = (2\pi)^{-4}(\alpha')^{-5/2}$, 
the free energy $\CE_4$ of the on-shell D4-branes wrapping $S^4$ at $u=u_c$ 
is given by $S_4|_{\hbox{\sevenrm on-shell}} =: -\int d\tau\, \CE_4$,  
\eqn\fourene{
\CE_4 = {\CN V_3 \over 3} u_c d \,.
}
We now consider the force balance condition between the baryonic D4-branes 
and the flavour D8-branes. 
The force of the D4-branes along the $u$ direction at $u=u_c$ 
is evaluated as 
\eqn\Cforcedfour{
f_4 = {1 \over \sqrt{g_{uu}}}\bigg|_{u=u_c}{d\CE_4 \over d u_c}
= {\CN V_3 \over 3R}d u_c^{3 \over 4}\sqrt{f(u_c)} \,.
}
Since the Legendre transformation of the D8-branes' action \CDeightDBI{} 
with respect to $a_\tau(u)$ becomes  
$$\eqalign{
{\tilde S_8} &= S_8^{\rm (DBI)} - \CN V_3 \int d\tau du\, d(u) a'_\tau(u) \cr
&= - \CN V_3 \int d\tau \int_{u_c}^\infty du\, u^4 \sqrt{
	\biggl( f(u)\bigl(x_4'(u)\bigr)^2+{1 \over u^3 f(u)}\biggr)
	\biggl( 1 + {\bigl(d(u)\bigr)^2 \over u^5}\biggr) } \,, 
}$$
we can read the tension of D8-branes at $u=u_c$ 
under the on-shell condition \CDeightsol{}, 
\eqn\Cforcedeight{
f_8 = {\CN V_3 \over R} u_c^{3 \over 4} \sqrt{u_c^5 + d^2 } \,,  
}
while the angle between the $u$ axis and the D8-branes is described as 
\eqn\Cangledeight{
\cos \theta = {\sqrt{g_{uu}} du \over \sqrt{g_{uu}du^2 + g_{44}dx_4^2}}\bigg|_{u=u_c} 
= \sqrt{1 -{f(u_0)(u_0^8 +u_0^3 d^2) \over f(u_c)(u_c^8 +u_c^3 d^2)} } \,.
}
Therefore the force yielded by the D8-branes along the $u$ direction 
is $f_8 \cos \theta$. 
Finally, from \Cforcedfour, \Cforcedeight\ and \Cangledeight, 
the force balance condition, 
$f_4 = f_8 \cos\theta$, leads to 
\eqn\Cforcebalance{
{1 \over 3}d = 
\sqrt{ {u_c^5 +d^2 \over f(u_c)} 
\biggl(1-{f(u_0)(u_0^8 +u_0^3 d^2) \over f(u_c)(u_c^8 +u_c^3 d^2)}\biggr) } \,.
}
One can show that $u_c$ is related to $\ell$, 
the separation between the D8-branes in the $x_4$ direction, by
\eqn\Cdeightsepa{
\ell = 2\int_{u_c}^\infty du\, x_4'(u) 
	= 2 \int_{u_c}^\infty du\, {1 \over u^{3 \over 2} f(u)}
	\biggl[{f(u)(u^8 +u^3 d^2) \over f(u_0)(u_0^8 +u_0^3d^2)} -1\biggr]^{-{1 \over 2}}\,.
}
Note that $\ell \leq \beta_4/2$. 
The $\ell =\beta_4/2$ case corresponds to the original Sakai-Sugimoto model.

\subsec{Chiral condensate}

In order to evaluate the chiral condensate, 
we need to compute the area of string instanton world-sheet \strWS, 
$$
S_\CO = {R^2 \over 2\pi\alpha'} \int_{-\ell/2}^{\ell/2} dx_4 \int_{u(x_4)}^\infty du 
= {R^2 \over \pi\alpha'} \int_0^{\ell/2} dx_4 \int_{u(x_4)}^\infty du \,.
$$
Since there is UV divergence in $S_\CO$, regularisation is necessary. 
In Ref.~\EV\ a UV cutoff parameter $u_\infty$ was introduced, 
however this regularisation neglects the contribution 
from the shape of D8-branes in the region of $u > u_\infty$.
Here we regularise $S_\CO$ 
by subtracting a density independent but infinite constant, 
$$
S_\infty := {R^2 \over \pi\alpha'} \int_0^{\ell/2} dx_4 \int_{u_{\rm KK}}^\infty du \,, 
$$
which is the area supported by two lines, $x_4= \pm \ell/2$. 
Then the regularized area $S_\CO^{\rm reg}$ is given  by 
\eqn\CregOWL{
S_\CO^{\rm reg} := S_\CO - S_\infty
= {R^2 \over 2\pi\alpha'} \ell u_{\rm KK} -{R^2 \over \pi\alpha'} \int_{u_c}^\infty du\, u x_4'(u) \,. 
}
Then the vacuum expectation value of open Wilson line operator is defined by 
\eqn\regulSinf{
\langle \CO \rangle_{\rm reg} := e^{-S_\CO^{\rm reg}} \,,
}
and it is identified as the chiral condensate 
$\langle {\bar \psi} \psi \rangle$. 

We shall compute it numerically, since  computing $S_\CO^{\rm reg}$ analytically is not easy.  
If we rescale the variables as 
\eqn\Cresc{
z := {u \over u_{\rm KK}} \,, \quad 
z_0 := {u_0 \over u_{\rm KK}} \,, \quad
z_c := {u_c \over u_{\rm KK}} \,, \quad 
\rho := {d \over u_{\rm KK}^{5/2}} \,, \quad 
{\tilde \ell} := \ell \sqrt{u_{\rm KK}} \,,
}
then one can exclude the appearance  of $u_{\rm KK}$ 
from the  numerical analysis. 
\Cforcebalance\ leads to 
\eqn\Cforcebalancereg{
Z_0(z_c,\rho) := (z_0^3 -1)(z_0^5 +\rho^2) = (z_c^3 -1)(z_c^5 +\rho^2)\biggl(1-{1 \over 9}\rho^2{1-z_c^{-3} \over z_c^5 +\rho^2}\biggr) \,,
}
and \Cdeightsepa\ can be rewritten as 
\eqn\Cdeightsepareg{
{\tilde \ell} = 2\int_{z_c}^\infty dz\, {z^{3/2} \over z^3-1}
	\biggl[{(z^3 -1)(z^5 +\rho^2) \over Z_0(z_c,\rho)} -1\biggr]^{-{1 \over 2}} \,.
}
${\tilde \ell}(z_c,\rho)$ is the monotonically decreasing function 
with respect to $z_c$ for any fixed $\rho$, 
and especially ${\tilde \ell}(1,\rho)$ is equal to 
$\sqrt{u_{\rm KK}} \beta_4/2 = 2\pi /3$, which means the antipodal position. 
The regularized area \CregOWL\ is 
\eqna\CGfunc
$$\eqalignno{
S_\CO^{\rm reg} &= {R^2 \over \pi\alpha'} \sqrt{u_{\rm KK}} G_{\rm conf} \,, &\CGfunc{a} \cr
G_{\rm conf} 
&= -\int_{z_c}^\infty dz\, {z^{3/2} \over z^2+z+1}
	\biggl[{(z^3 -1)(z^5 +\rho^2) \over Z_0(z_c,\rho)} -1\biggr]^{-{1 \over 2}} \,. &\CGfunc{b} 
}$$

\bigskip\vbox{
\centerline{\epsfbox{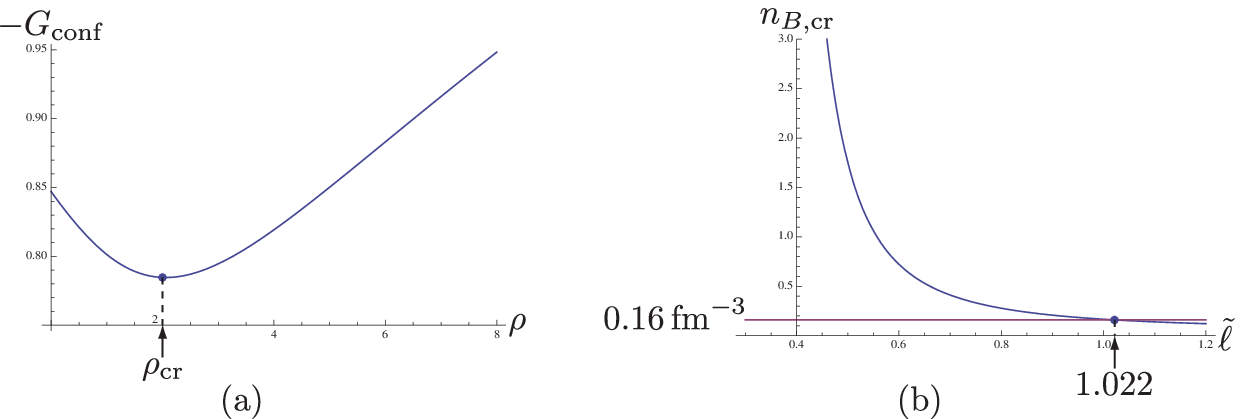}} 
\centerline{\vbox{\offinterlineskip 
\halign{ \strut# & #\hfil \cr
\fig\figCCconf{} & (a) The numerical plot of $-G_{\rm conf}(\rho)$ with ${\tilde \ell}=1/2$ fixed. \cr
& (b) The critical density $n_{B,{\rm cr}}({\tilde \ell})$. \cr
}}}}
\bigskip
From now on we set   ${\tilde \ell} = 1/2$ for definiteness of our model. 
The numerical plot of $-G_{\rm conf}(\rho)$, the regularized area,  is depicted by \figCCconf a. 
From this plot, we can find that $\langle \CO \rangle_{\rm reg}$ becomes minimum 
at $\rho_{\rm cr} \approx 2.058$. 
Setting  $N_c = 3$ and $N_f = 2$, we evaluate 
the baryon number density at $\rho = \rho_{\rm cr}$ 
in terms of \BNdensity\ and \dnfour, namely, 
\eqn\numdens{
n_B = {N_f M_{\rm KK}^3\lambda^2 \over 1458\pi^4} \rho \,. 
}
Here we are interested in the low density region 
where the chiral condensate ($\sim \langle \CO \rangle_{\rm reg}$) 
is expected to decreases. 
We fix $M_{\rm KK}$ and $\lambda$ by the use of 
the generalized Sakai-Sugimoto model at zero density 
\refs{\PSM,\SeSo}. 
Using the experimental values of 
the mass of $\rho$ meson, $m_\rho \approx 776 \,{\rm MeV}$, 
and the pion decay constant, $f_\pi \approx 93 \,{\rm MeV}$,
one can obtain 
$$
M_{\rm KK} \approx 496 \,{\rm MeV}\,, \quad 
\lambda \approx 61.7 \,.
$$
Note that  these values  are for  the non-antipodal model with ${\tilde \ell}=1/2$, 
 and  different from those in the usual antipodal Sakai-Sugimoto model. 
Then the number density \numdens\ at $\rho_{\rm cr} \approx 2.058$ 
is evaluated to be 
$$
n_{B,{\rm cr}} \approx 1.76 \, {\rm fm}^{-3} \,.
$$
In terms of the normal nuclear density, $n_0 = 0.16\, {\rm fm}^{-3}$, 
we obtain the ratio, $n_{B,{\rm cr}} / n_0 \approx 11$, 
therefore $n_{B,{\rm cr}}$ is sufficiently dense. 

For any other ${\tilde \ell}$, the behaviour of $\langle \CO \rangle_{\rm reg}$ 
is schematically same as the case of ${\tilde \ell} = 1/2$, 
that is to say, the regularized area  has minimum at $\rho_{\rm cr}$ 
as a function  of density. 
Once we choose a model by fixing ${\tilde \ell}$, 
then $\rho_{\rm cr}$ is determined. 
$n_{B,{\rm cr}}$ calculated from $\rho_{\rm cr}$ through \numdens\ is 
the function of ${\tilde \ell}$, which is depicted by \figCCconf b. 
$n_{B,{\rm cr}}$ becomes equal to the normal baryon density 
at ${\tilde \ell} \approx 1.022$.

\newsec{Deconfined phase}

In this section we concentrate on the deconfined phase of 
the generalised Sakai-Sugimoto model with a baryon density. 
This phase is similar to the holographic dual of NJL model \AHJK.  
Since the fifth direction $(x_4)$ is not compact in the holographic NJL model, 
the Kaluza-Klein scale $M_{\rm KK}$ does not appear. 
On the other hand, in our case the $x_4$ direction 
is compactified, so that we can compare our results with experimental values 
by calibrating $M_{\rm KK}$. 
The open Wilson line in the holographic NJL model with a baryon density 
has been studied by Ref.~\EV, whose analysis is quite similar to ours 
in the generalised Sakai-Sugimoto model. 
However we shall use the different regularisation scheme 
from Ref.~\EV\ in order to include more accurate UV behaviour, 
and compute physical values by the use 
of the calibrated $M_{\rm KK}$ and $\lambda$. 

\subsec{D-brane configuration and force balance condition}

We turn on the $U(1)$ gauge field defined by \uoneat. 
Then the DBI action of $N_f$ flavour D8-branes in the background \Dmetric{} 
corresponding to the deconfined phase is 
\eqna\DDeightDBI
$$\eqalignno{
&S_8^{\rm (DBI)} = -\CN V_3 \beta_\tau \int du\, \CL[x_4', a'_\tau] \,, &\DDeightDBI{a} \cr
&\CL[x_4', a'_\tau] = u^4 \sqrt{f_T(u)\bigl(x_4'(u)\bigr)^2 -{1 \over u^3}\bigl(a_\tau'(u)\bigr)^2 +{1 \over u^3} } \,. &\DDeightDBI{b}
}$$
Note that the last term in the square root of \DDeightDBI{b} is different from that of \CDeightDBI{b} by the absence of $1/f$ factor. 
We recall that $\beta_\tau$ is the period of $\tau$ direction given by \Dtper, 
therefore the temperature in the deconfined phase is denoted by 
$T= \beta_\tau^{-1} = (3/4\pi)\sqrt{u_T}$. 
In the same way as Section 3, we define the electric displacement field 
by $d(u)\equiv -\delta \CL /\delta a_\tau'$ ({\it cf.}~\Celedis), 
\eqn\Deledis{
d(u) = u a_\tau'(u)\biggl[f_T(u)\bigl(x_4'(u)\bigr)^2 -{1 \over u^3}\bigl(a_\tau'(u)\bigr)^2 +{1 \over u^3}\biggr]^{-{1 \over 2}} \,.
}
Then the equations of motion for $x_4(u)$ and $a_\tau(u)$ are written down as 
\eqna\Deom
$$\eqalignno{
0&= {d \over du}\Biggl[u^4 f_T(u) x_4'(u) 
	\biggl(1 + {\bigl(d(u)\bigr)^2 \over u^5}\biggr)^{1 \over 2} 
	\biggl( f_T(u) \bigl(x_4'(u)\bigr)^2 +{1 \over u^3 f_T(u)} \biggr)^{-{1 \over 2}} 
\Biggr] \,, &\Deom{a} \cr
0&=d'(u) \,. &\Deom{b} \cr
}$$

In the deconfined phase, the following constant solution is allowed: 
\eqn\Dparasol{
x_4(u) = \pm {1 \over 2} \ell \,, \quad d(u) = d \,.
}
This describes the parallel D8-branes and anti-D8-branes 
whose separation is  $\ell$  (see \figphases c). 
Therefore \Dparasol\ implies that the chiral symmetry is restored. 

One can also find a non-trivial solution of \Deom{}:  
\eqna\DDeightsol
$$\eqalignno{
\bigl(x_4'(u)\bigr)^2 &= {1 \over u^3 f_T(u)}\biggl[{f_T(u)(u^8 +u^3 d^2) \over f_T(u_0)(u_0^8 +u_0^3 d^2)} -1\biggr]^{-1} \,, &\DDeightsol{a} \cr
d(u) &= d \,. &\DDeightsol{b}
}$$
This solution corresponds to the V-shape (or U-shape) D8-branes (\figphases b).
From \DDeightsol{a} we can calculate the separation between the D8-branes 
at $u = \infty$, 
\eqn\Ddeightsepa{
\ell = 2\int_{u_c}^\infty du\, x_4'(u) 
= 2\int_{u_c}^\infty du\, {1 \over u^{3/2} \sqrt{f_T(u)}}\biggl[{f_T(u)(u^8 +u^3 d^2) \over f_T(u_0)(u_0^8 +u_0^3 d^2)} -1\biggr]^{-{1 \over 2}} \,.
}

The source term comes from the Chern-Simons action of the D8-branes 
as we explained in Section 3. 
We assume the uniform distribution of baryons in $\Bbb{R}^3$ 
whose density is denoted by $n_4$. 
By integrating the equation of motion for $a_\tau(u)$ which is derived 
from the DBI action and additionally the source term, 
the constants $d$ and $n_4$ are related by \dnfour\ again in the deconfined phase. 
The action of $N_4 (= n_4 V_3)$ baryonic D4-branes wrapping $S^4$ at $u=u_c$ 
is calculated as 
\eqn\Ddfouract{
S_4 = -{\CN V_3 \beta_\tau \over 3}d u_c \sqrt{f_T(u_c)} \,,
}
and the force generated by these D4-branes along the $u$ direction is 
evaluated as 
\eqn\Dforcedfour{
f_4 = {\CN V_3 \beta_\tau d \over 3R} u_c^{3 \over 4}\Biggl[1+{1 \over 2}\biggl({u_T \over u_c}\biggr)^3\Biggr] 
= {\CN V_3 \beta_\tau \over 2R} d u_c^{3 \over 4} \biggl(1-{1 \over 3}f_T(u_c) \biggr) \,.
}
We now compute the force from the flavour D8-branes. 
From the Legendre transformed action, 
$$
{\tilde S}_8 = -\CN V_3 \beta_\tau \int du\, u^4 \sqrt{
	\biggl(f_T(u) \bigl(x_4'(u)\bigr)^2 +{1 \over u^3}\biggr)
	\biggl(1 + {\bigl(d(u)\bigr)^2 \over u^5}\biggr)} \,,
$$
the tension of the flavour D8-branes is computed to be 
\eqn\Dforcedeight{
f_8 = {\CN V_3 \beta_\tau \over R}u_c^{3 \over 4}
	\sqrt{f_T(u_c) \bigl(u_c^5 + d^2 \bigr)} \,.
}
Since the angle between the D8-branes and the $u$ axis is 
\eqn\Dangledeight{
\cos \theta 
= \sqrt{ 1 -{f_T(u_0)(u_0^8 +u_0^3 d^2) \over f_T(u_c)(u_c^8 +u_c^3 d^2)} } \,,
}
the force coming from the D8-brane along the $u$ direction becomes 
$f_8 \cos \theta$.
From \Dforcedfour, \Dforcedeight\ and \Dangledeight, 
we can obtain the force balance condition, $f_4 = f_8 \cos \theta$, that is,  
\eqn\Dforcebalance{
{d \over 2} \biggl(1-{1 \over 3}f_T(u_c)\biggr) = \sqrt{ f_T(u_c) (u_c^5 +d^2)\biggl(1 -{f_T(u_0)(u_0^8 +u_0^3 d^2) \over f_T(u_c)(u_c^8 +u_c^3 d^2)}\biggr) } \,.
}

\subsec{Chiral condensate}

We redefine the coordinates by 
\eqn\Dresc{
z := {u \over u_T} \,, \quad 
z_0 := {u_0 \over u_T} \,, \quad
z_c := {u_c \over u_T} \,, \quad 
\rho := {d \over u_T^{5/2}} \,, \quad 
{\tilde \ell} := \ell \sqrt{u_T} \,,
}
in the similar way to \Cresc. 
Then \Ddeightsepa\ leads to 
\eqn\Dforcebalancereg{
Z_0(z_c,\rho) := (z_0^3 -1)(z_0^5 +\rho^2) = (z_c^3 -1)(z_c^5 +\rho^2) 
\biggl[ 1  -{1 \over 36}\rho^2{(1+2z_c^3)^2 \over z_c^3 (z_c^3 -1)(z_c^5 +\rho^2)} \biggr] \,,
}
and one can rewrite \Dforcebalance\ as 
\eqn\Ddeightsepareg{
{\tilde \ell}= 2\int_{z_c}^\infty dz\, {1 \over \sqrt{z^3 -1}}\biggl[{(z^3 -1)(z^5 +\rho^2) \over Z_0(z_c,\rho)} -1\biggr]^{-{1 \over 2}} \,. 
}
\bigskip\vbox{
\centerline{\epsfbox{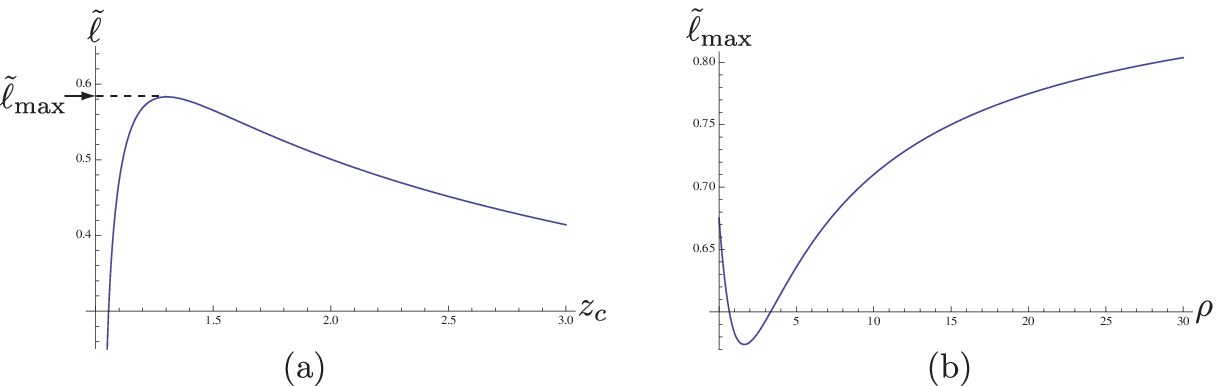}} 
\centerline{\vbox{\offinterlineskip 
\halign{ \strut# & #\hfil \cr
\fig\figlmax{} & (a) The plot of ${\tilde \ell}$ as function of $z_c$ 
	with $\rho = 1$. \cr
& (b) The $\rho$ dependence of ${\tilde \ell}_{\rm max}(\rho)$. \cr
}}}}
\bigskip\nobreak\noindent
\Ddeightsepareg\ includes three variables, ${\tilde \ell}$, $\rho$ and $z_c$. 
As Ref.~\BLL\ has pointed out,\foot{Ref.~\BLL\ considered 
in the variables $\ell$ and $d$ with fixed temperature. 
However we here use the rescaled variables \Dresc, such that 
the temperature factor $u_T$ is excluded from the following numerical analyses. 
} 
for any combination of ${\tilde \ell}$ and $\rho$, 
$z_c$ does not necessarily exist, 
because the right hand side of \Ddeightsepareg\ with a fixed $\rho$ 
is the function of $z_c$, which has a maximum, ${\tilde \ell}_{\rm max}$ 
(see \figlmax a). 
This is a remarkable feature different from the confined phase. 
By changing $\rho$, we depict the $\rho$ dependence of ${\tilde \ell}_{\rm max}$
in \figlmax b. 
For  ${\tilde \ell} > {\tilde \ell}_{\rm max}$, 
there is no solution of \Ddeightsepareg, that is, the V-shape solution 
does not exist, but there is only the parallel solution \Dparasol. 
On the other hand, for the opposite case, as one can see from \figlmax a,  there are two solutions in $z_c$ 
of \Ddeightsepareg\  for the given value of  ${\tilde \ell}\, (< {\tilde \ell}_{\rm max})$ 
and $\rho$. 
These two solutions describe different configurations of V-shape D8-branes. 
{\it By comparing the free energies for these two configurations, one can clarify that the larger $z_c$ is favoured. }
Therefore we hereafter use only the larger $z_c$. 

Let us evaluate the open Wilson line under the constraints given by \Dforcebalancereg\ and \Ddeightsepareg. 
The on-shell world-sheet action of open Wilson line \strWS\ in the 
background \Dmetric{} is 
$$
S_\CO = {R^2 \over \pi\alpha'} \int_{0}^{\ell/2} dx_4 \int_{u(x_4)}^\infty du\, {1 \over \sqrt{f_T(u)}} \,.
$$
Since this integration diverges by the contribution of UV region, 
we have to regularise it in the similar manner done in the previous section, 
that is, we subtract the infinite constant defined by 
$$
S_\infty := {R^2 \over \pi\alpha'} \int_{0}^{\ell/2} dx_4 \int_{u_T}^\infty du\, {1 \over \sqrt{f_T(u)}} \,.
$$ 
Since this area is supported by the two parallel lines \Dparasol, 
$S_\infty$ is related to ``chiral condensate'' 
in the chiral symmetry restoration. 
We shall give more comments about this issue in Section 5. 
Then the regularised area becomes 
\eqna\DGdeconf
$$\eqalignno{
S_\CO^{\rm reg} &= S_\CO -S_\infty  = {R^2 \over \pi\alpha'} \sqrt{u_T}G_{\rm deconf} \,, &\DGdeconf{a} \cr
G_{\rm deconf} 
&=-\int_{z_c}^\infty dz\, \biggl(\int_1^z{d\zeta \over \sqrt{1-\zeta^{-3}}}\biggr){1 \over \sqrt{z^3 -1}}\biggl[{(z^3 -1)(z^5 +\rho^2) \over Z_0(z_c,\rho)} -1\biggr]^{-{1 \over 2}} \,, &\DGdeconf{b}
}$$
where we used \regulSinf\ and \Dforcebalancereg.

\bigskip\vbox{
\centerline{\epsfbox{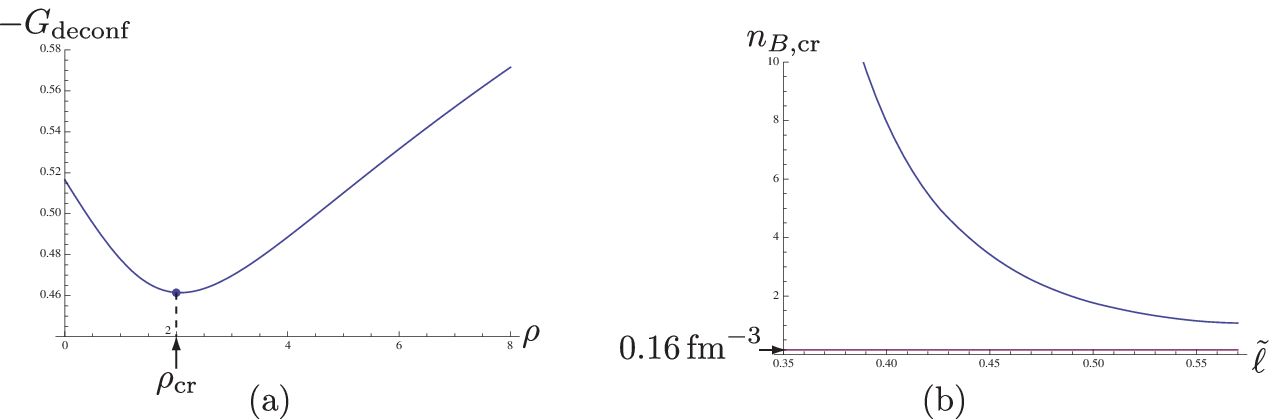}} 
\centerline{\vbox{\offinterlineskip 
\halign{ \strut# & #\hfil \cr
\fig\figCCdeconf{} & (a) The numerical plot of $-G_{\rm deconf}(\rho)$ 
	with ${\tilde \ell}=1/2$ fixed. \cr
& (b) The numerical plot of $n_{B,cr}({\tilde \ell})$. \cr
}}}}
\bigskip 
Let us set ${\tilde \ell} = 1/2$ for example. 
The numerical plot of $-G_{\rm deconf}$ is shown in \figCCdeconf a. 
It implies that in the deconfined phase also $\langle \CO \rangle_{\rm reg}$ 
firstly decreases and then increases with respect to the density $\rho$. 
$\langle \CO \rangle_{\rm reg}$ has a minimum value at 
\eqn\Drhocr{
\rho_{\rm cr} \approx 2.073 \,. 
} 
This is interpreted by the use of \numdens\ to the dimensionful value, 
$$
n_{B,{\rm cr}} = 1.771\, {\rm fm}^{-3} \approx 11 n_0 \,, 
$$
where we assumed $u_{\rm KK} = u_T$ for simplicity.\foot{If we consider 
arbitrary $u_{\rm KK}$ and $u_T$, we have to take care on 
the difference of scalings (see \Cresc\ and \Dresc), 
${\tilde \ell}_{\rm (confine)} 
= \sqrt{u_{\rm KK}/u_T}\,{\tilde \ell}_{\rm (deconfine)}\, (= \ell)$,  
in calibrating $M_{\rm KK}$ and $\lambda$. 
The decreasing behaviour of $n_{B,{\rm cr}}$ schematically does not change. 
}

For arbitrary ${\tilde \ell}$, 
$\rho_{\rm cr}$ is given by a function of ${\tilde \ell}$. 
In \figCCdeconf b, we plot $n_{B,{\rm cr}}({\tilde \ell})$ 
under the assumption of $u_{\rm KK} = u_T$. The chiral condensate monotonically decreases 
in the region $n_B < n_{B,{\rm cr}}$. 
A remarkable feature in \figCCdeconf b\ is 
that $n_{B,{\rm cr}}$ is sufficiently larger than the normal nuclear density.

\subsec{Can the chiral symmetry be restored by the density effect?}

We have to clarify which solution of the D8-branes is favoured 
with respect to the baryon density, the parallel configuration \Dparasol\ 
or the V-shape one \DDeightsol{}. 
For this purpose we compare the free energies of these two configurations 
({\it cf.}~Ref.~\ASY\ in zero density).  
The energy of the V-shape is given by 
the flavour D8-branes and the baryon vertex of D4-branes. 
On the other hand, the energy of the parallel configuration consists of 
only the D8-branes, 
because the D4-branes of baryons disappear into the black hole. 
The discrepancy of the free energies between those configurations is 
$$
{\tilde S}_8^{\rm (V)} + S_4^{\rm (V)} - {\tilde S}_8^{\rm (parallel)} 
=: -\CN V_3 \beta_\tau u_T^{7 \over 2} \Delta \CE(\rho;{\tilde \ell}) \,,
$$
where 
$$\eqalign{
\Delta \CE(\rho;{\tilde \ell}) =& \int_{z_c}^\infty dz\, \sqrt{z^5+\rho^2}\biggl[ \biggl(1-{(z_0^3-1)(z_0^5 +\rho^2) \over (z^3-1)(z^5 +\rho^2)}\biggr)^{-{1 \over 2}} -1 \biggr] 
-\int_1^{z_c} dz\, \sqrt{z^5+\rho^2} \cr
&+{1 \over 3}\rho z_c \sqrt{1-z_c^{-3}}  \,.
}$$
Note that there are the overall minus signs in the actions. 
Therefore, if $\Delta \CE$ is positive (negative), 
then the parallel configuration (the V-shape configuration) is favoured. 

\bigskip\vbox{
\centerline{\epsfbox{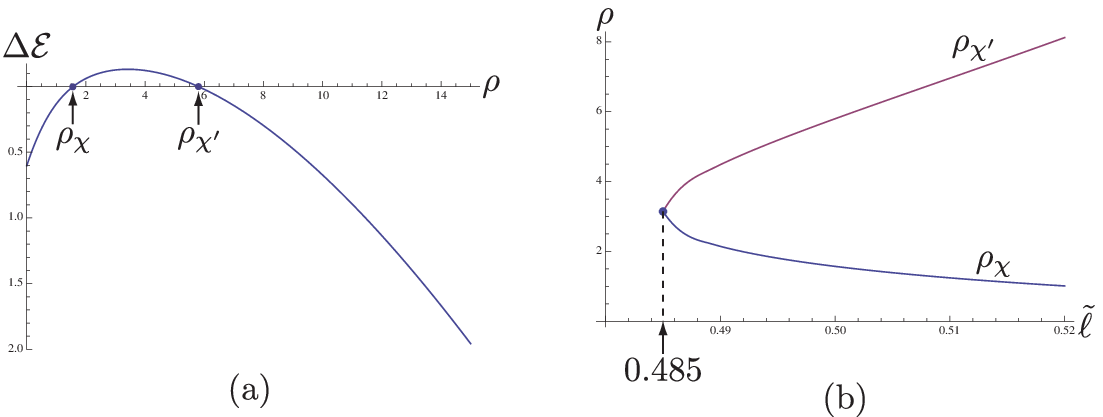}} 
\centerline{\vbox{\offinterlineskip 
\halign{ \strut# & #\hfil \cr
\fig\figchisr{} & (a) The plot of $\Delta\CE(\rho;{\tilde \ell}=1/2)$, 
whose zeros are $\rho_\chi$ and $\rho_{\chi'}$. \cr
& (b) The plots of $\rho_\chi({\tilde \ell})$ and $\rho_{\chi'}({\tilde \ell})$. 
Their crossing point is approximately equal \cr
&\phantom{(b)} to $(0.485,3.144)$. \cr
}}}}
\bigskip 
For example, fixing ${\tilde \ell}=1/2$ again, 
we compute $\Delta \CE(\rho;{\tilde \ell}=1/2)$. 
The   plot for  $\Delta \CE(\rho;{\tilde \ell}=1/2)$ 
  given  in \figchisr a shows 
that $\Delta \CE$ has two zeros, namely, 
the smaller zero, $\rho_\chi \approx 1.571$, and 
the larger one, $\rho_{\chi'} \approx 5.798$. 
In the region, $\rho < \rho_\chi$ and $\rho > \rho_{\chi'}$, 
the chiral symmetry is broken, 
while in the region, $\rho_\chi < \rho < \rho_{\chi'}$, 
the chiral symmetry is restored. 
In the ${\tilde \ell}=1/2$ model, 
$\rho_\chi$ is smaller than $\rho_{\rm cr}$ (see \Drhocr). 
Therefore the chiral symmetry is restored 
before the chiral condensate $\langle \CO \rangle_{\rm reg}$ starts to increase. 
Since one can regard $S_\CO$ as $S_\infty$, {\it i.e.}~$S_\CO^{\rm reg} = 0$, 
in the parallel D8-branes solution, 
the chiral condensate becomes $\langle \CO \rangle_{\rm reg}=1$ 
in the chiral symmetry restored phase. 

Let us consider an arbitrary ${\tilde \ell}$. 
The chiral symmetry restoration/breaking points, $\rho_\chi$ and $\rho_{\chi'}$, 
do not necessarily exist for any ${\tilde \ell}$. 
The ${\tilde \ell}$ dependences of $\rho_\chi$ and $\rho_{\chi'}$ 
are depicted in \figchisr b, which says that 
$\rho_\chi$ and $\rho_{\chi'}$ do not exist when ${\tilde \ell} < 0.485$. 
In other words, the model with ${\tilde \ell} < 0.485$ does not have 
the  phase  of chiral symmetry restoration. 
On the other hand, in the model with ${\tilde \ell} > 0.485$, 
the chiral symmetry is restored only in the window region:  $\rho_\chi <\rho < \rho_{\chi'}$.
\bigskip\vbox{
\centerline{\epsfbox{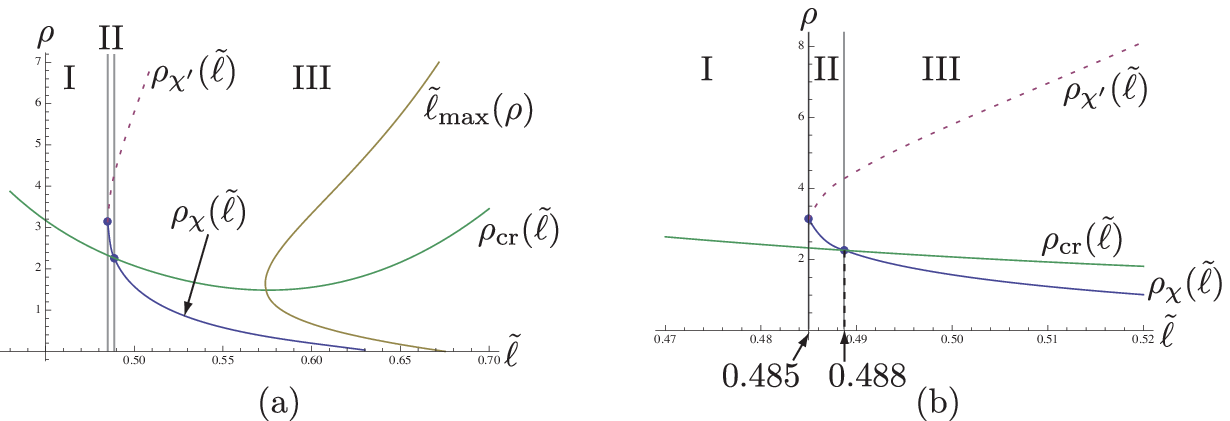}} 
\centerline{\vbox{\offinterlineskip 
\halign{ \strut# & #\hfil \cr
\fig\figchisdiag{} & (a) The plots of $\rho_{\rm cr}({\tilde \ell})$, $\rho_\chi({\tilde \ell})$, $\rho_{\chi'}({\tilde \ell})$ and ${\tilde \ell}_{\rm max}(\rho)$. \cr
& (b) The magnified plot of (a) around the crossing points of $\rho_{\rm cr}$, 
	$\rho_\chi$ and $\rho_{\chi'}$. \cr
}}}}
\bigskip\nobreak\noindent

Combining $\rho_{\rm cr}({\tilde \ell})$, $\rho_\chi({\tilde \ell})$, 
$\rho_{\chi'}({\tilde \ell})$ and ${\tilde \ell}_{\rm max}(\rho)$
that we have   calculated so far, 
we obtain \figchisdiag. 
In the region of ${\tilde \ell} > {\tilde \ell}_{\rm max}$, 
the chiral symmetry is restored, because there does not 
exist the V-shape solution but the parallel one. 
Also in the region of ${\tilde \ell} \gtrsim 0.485$ and 
$\rho_\chi < \rho < \rho_{\chi'}$, the chiral symmetry is restored, 
because the free energy of the parallel solution is smaller than 
that of the V-shape solution. 
In the other region, the chiral symmetry is broken. 
From \figchisdiag, we can find the following three types of models 
on the chiral condensate $\langle \CO \rangle_{\rm reg}$:
\bigskip\vbox{
\centerline{\epsfbox{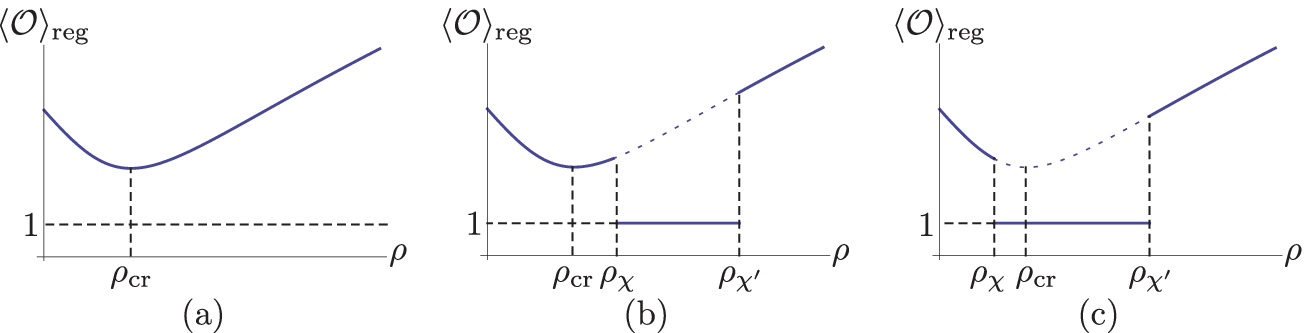}} 
\centerline{\vbox{\offinterlineskip 
\halign{ \strut# & #\hfil \cr
\fig\figccphases{} & The (regularised) chiral condensate 
	in (a) type I, (b) type II, (c) type III. \cr
}}}}
\bigskip\nobreak\noindent
\item{I.} ${\tilde \ell} < 0.485$ \hfil\break 
Since the chiral symmetry is always broken, in other words, 
the chiral symmetry is never restored. 
$\langle \CO \rangle_{\rm reg}$ firstly decreases and then starts to increase. 
This is depicted by \figccphases a.
\medskip
\item{II.} $0.485 < {\tilde \ell} < 0.488$ \hfil\break 
In this type, $\rho_{\rm cr} < \rho_\chi < \rho_{\chi'}$. 
Therefore the chiral condensate decreases in $\rho < \rho_{\rm cr}$ 
and increases in $\rho_{\rm cr}< \rho <\rho_\chi$. 
In $\rho_\chi< \rho < \rho_{\chi'}$ the chiral symmetry is restored, 
so that $\langle \CO \rangle_{\rm reg}$ is equal to one. 
When $\rho$ is larger than $\rho_{\chi'}$, the chiral symmetry is broken again. 
The behaviour of $\langle \CO \rangle_{\rm reg}$ is depicted by \figccphases b. 
\medskip
\item{III.} ${\tilde \ell} > 0.488$ \hfil\break 
Since $\rho_\chi$ is smaller than $\rho_{\rm cr}$, 
before $\langle \CO \rangle_{\rm reg}$ goes up, 
the chiral symmetry is restored, 
namely, $\langle \CO \rangle_{\rm reg}$ becomes one. 
Then in $\rho > \rho_{\chi'}$ the chiral symmetry is broken again 
and $\langle \CO \rangle_{\rm reg}$ increases. 
This behaviour is shown in \figccphases c and this case corresponds to the expected result from the 
field theory intuition. 
\medskip
\noindent
In models with any ${\tilde \ell}$, 
$\langle \CO \rangle_{\rm reg}$ almost linearly decreases in low density, 
and increases at very high density. 
The type II and type III have the transition 
to the chiral symmetry restored phase at $\rho = \rho_\chi$.

\newsec{Conclusions and discussions}

We have studied the chiral condensate given by the open Wilson line 
in the generalised Sakai-Sugimoto model with the baryon density. 
In order to exclude $u_{\rm KK}$ and $u_T$ dependences from 
numerical analyses, we introduced the rescaled variables, \Cresc\ and \Dresc.
In both of the confined and deconfined phases, 
the chiral condensate firstly decreases and then increases 
with respect to the density $\rho$ (equivalently $n_B$). 
We have calculated the critical density $n_{B,{\rm cr}}$ 
at the turning point, which depends on ${\tilde \ell}$. 
As a result, $n_{B,{\rm cr}}$ is the decreasing function of ${\tilde \ell}$. 
For instance, in the model of ${\tilde \ell}=1/2$, 
$n_{B,{\rm cr}}$ is about eleven times as large as 
the normal nuclear density $n_0$. 
Therefore the range, $n_B < n_{B,{\rm cr}}$, 
in which the chiral condensate decreases, is sufficiently large, 
and this decreasing behaviour agrees with our intuition from ordinary QCD. 
However the increasing behaviour in very high density is 
different from the expectation from QCD. 

There is not chiral symmetry restoration in the confined phase. 
On the other hand, in the deconfined phase, 
there is the region of ${\tilde \ell}$ and $\rho$ 
in which the chiral symmetry is restored, 
so that we have found the three types of behaviour 
of $\langle \CO \rangle_{\rm reg}$ with respect to ${\tilde \ell}$ (see \figccphases). 
In any types, the decreasing behaviour of $\langle \CO \rangle_{\rm reg}$ 
in low density
is consistent with the results which we expect from the ordinary QCD. 
In the models of type II and III, there is a transition 
to the chiral symmetry restoration at $\rho = \rho_\chi$, 
especially the models in the type III, {\it i.e.}~${\tilde \ell} > 0.488$, 
are in good agreement with QCD. 
However the chiral symmetry restoration occurs 
before $\langle \CO \rangle_{\rm reg}$ reaches one, 
that is to say, this transition is in first order. 

In all types the chiral condensate grows in very high density, 
and this behaviour disagrees with our intuition from QCD. 
A possible problem in the high density is a backreaction of 
the baryonic D4-branes. 
Since the high density means the large number of baryonic D4-branes, 
the background geometry should be modified by these D4-branes. 
Ref.~\RSZ\ has mentioned about such kind of modification. 
The most honest way to deal with dense baryons 
in the generalised Sakai-Sugimoto model is 
to find a classical solution in supergravity with $N_c$ colour D4-branes 
and $N_4$ baryonic D4-branes and to deal with the flavour D8-branes as a probe, 
but this would be difficult. 

As we studied so far, the chiral condensate $\langle \CO \rangle$ itself
defined by the open Wilson line is not appropriate for 
the order parameter of chiral symmetry. 
However the positivity of $-S_\CO^{\rm reg}$ by definition, namely, 
$\langle \CO \rangle_{\rm reg} \geq 1$, 
is a good tendency to the order parameter. 
Therefore we  suggest  
the chiral symmetry order parameter $\chi$ defined by 
\eqn\chiOparam{
\chi := \langle \CO \rangle_{\rm reg} -1 
}
so that $\chi$ vanishes  only when the chiral symmetry is restored: 
$S_\CO$ is equal to $S_\infty$ in the parallel D8-branes 
and anti-D8-branes configuration. 
When the chiral symmetry is broken, 
$\chi$ is always positive (see also \figccphases).

\bigbreak\bigskip\bigskip\centerline{{\bf Acknowledgment}}\nobreak
SS is grateful to Center for Quantum Spacetime (CQUeST) in Sogang University 
and Institut des Haute {\' E}tudes Scientifiques (IH{\' E}S) for hospitality. 
The work of SJS was supported by Mid-career Researcher Program through NRF grant No.~2010-0008456. 
It was also partly supported by the National Research Foundation of Korea (NRF) 
grant through the SRC program CQUeST with grant number 2005-0049409.

\listrefs 

\bye